\documentclass[11pt,onecolumn]{article}
\usepackage[dvips]{graphics}
\usepackage[dvips]{color}
\usepackage{amssymb}
\usepackage{amsmath}
\usepackage{amsthm}
\usepackage[dvips]{geometry}
\usepackage{setspace}
\theoremstyle{plain}
\newtheorem{theorem}{Theorem}
\newtheorem{lemma}{Lemma}

\def\limp{\mathop{\rm lim\vphantom{p}}}%

\newcommand{\iid}{\text{i.i.d.}}

\newcommand{\eps}{\epsilon}

\newcommand{\Verdu}{Verd\'u}
\newcommand{\Kavcic}{Kav\v{c}i\'c}

\begin{document}

\title{A Coding Theorem for a Class of Stationary 
Channels with Feedback}
\author{Young-Han Kim\\
University of California, San Diego}
\date{\today}
\maketitle

\begin{abstract}
A coding theorem is proved for a class of stationary channels with
feedback in which the output $Y_n = f(X_{n-m}^n, Z_{n-m}^n)$ is the
function of the current and past $m$ symbols from the channel input
$X_n$ and the stationary ergodic channel noise $Z_n$.  In particular,
it is shown that the feedback capacity is equal to
\[
\limp_{n\to\infty} \sup_{p(x^n||y^{n-1})} 
\frac{1}{n} I(X^n \to Y^n),
\]
where $I(X^n \to Y^n) = \sum_{i=1}^n I(X^i; Y_i|Y^{i-1})$ denotes the
Massey directed information from the 
channel input to the output, and
the supremum is taken over all causally conditioned distributions
$p(x^n||y^{n-1}) = \prod_{i=1}^n p(x_i|x^{i-1},y^{i-1})$.  The main
ideas of the proof are the Shannon strategy for coding with side
information and a new elementary coding technique for the given
channel model \emph{without feedback}, which is in a sense dual to
Gallager's lossy coding of stationary ergodic sources.  A similar
approach gives a simple alternative proof of coding theorems for
finite state channels by Yang--\Kavcic--Tatikonda, Chen--Berger, and
Permuter--Weissman--Goldsmith.
\end{abstract}

\section{Introduction}
\label{sec:intro}
Shannon~\cite{Shannon1948} showed that the capacity $C$ of a
memoryless channel $(\mathcal{X}, p(y|x), \mathcal{Y})$, operationally
defined as supremum of all achievable rates~\cite[Section
7.5]{Cover--Thomas2006}, is characterized by
\begin{equation}
\label{eq:dmc-capacity}
C = \sup_{p(x)} I(X;Y).
\end{equation}
When the channel has memory but still maintains certain ergodic
properties, then \eqref{eq:dmc-capacity} can be extended to the
following multi-letter expression:
\begin{equation}
\label{eq:multi}
C = \limp_{n\to\infty} \sup_{p(x^n)} \frac{1}{n} I(X^n; Y^n).
\end{equation}
For example, Dobrushin~\cite{Dobrushin1963} showed that the capacity
formula \eqref{eq:multi} holds if the channel is \emph{information
stable}; see also Pinsker~\cite{Pinsker1964}. Further extensions and
refinements of \eqref{eq:multi} with more general capacity formulas
abound in the literature.  For stationary channels, readers are
referred to Gray and Ornstein~\cite{Gray--Ornstein1979},
Kieffer~\cite{Kieffer1981a}, and the references therein.  A general
formula for the capacity is given by \Verdu{} and
Han~\cite{Verdu--Han1994} for arbitrary nonstationary channels that
can be represented through a sequence of $n$-dimensional conditional
distributions (even without any consistency requirement); see also
Han~\cite{Han2003}.

For memoryless channels \emph{with feedback}, it was again
Shannon~\cite{Shannon1956} who showed that feedback does not increase
the capacity and hence that the feedback capacity is given by
\begin{equation}
\label{eq:dmc-fb}
C_\textsl{FB} = C = \sup_{p(x)} I(X;Y).
\end{equation}
As in the case of nonfeedback capacity~\eqref{eq:multi}, the question
arises how to extend the feedback capacity formula \eqref{eq:dmc-fb}
to channels with memory.  The most natural candidate is the following
multi-letter expression with \emph{directed information} introduced by
Massey~\cite{Massey1990} in place of the usual mutual information in
\eqref{eq:multi}:
\begin{align}
\label{eq:multi-fb}
C_\textsl{FB} &= \limp_{n\to\infty} \sup_{p(x^n||y^{n-1})}
\frac{1}{n}I(X^n \to Y^n) \\ 
&= \limp_{n\to\infty} \sup_{p(x^n||y^{n-1})} \frac{1}{n}\sum_{i=1}^n
I(X^i; Y_i|Y^{i-1}), \nonumber
\end{align}
where the supremum is taken over all $n$-dimensional \emph{causally
conditioned} probabilities
\begin{align*}
p(x^n||y^{n-1}) &= \prod_{i=1}^n p(x_i|x^{i-1}, y^{i-1})\\
&=
p(x_1) p(x_2|x_1,y_1)\cdots p(x_n|x_1,\ldots, x_{n-1}, y_1, \ldots, y_{n-1}).
\end{align*}

The main goal of this paper is to establish the validity of the
feedback capacity formula~\eqref{eq:multi-fb} for a reasonably general
class of channels with memory, in the simplest manner.

Massey~\cite{Massey1990} introduced the mathematical notion of
{directed information}
\[
I(X^n \to Y^n) = \sum_{i=1}^n I(X^i; Y_i|Y^{i-1}),
\]
and established its operational meaning by showing that the feedback
capacity is upper bounded by the maximum normalized directed
information, which can be in general tighter than the usual mutual
information.  He also showed that \eqref{eq:multi-fb} reduces to
\eqref{eq:dmc-fb} if the channel is memoryless, and to
\eqref{eq:multi} if the channel is used without feedback.
Kramer~\cite{Kramer1998, Kramer2003} streamlined the notion of
directed information further and explored many interesting properties;
see also Massey and Massey~\cite{Massey--Massey2005}.

For channels with certain structures, the validity of the feedback
capacity formula \eqref{eq:multi-fb} has been established implicitly.
For example, Cover and Pombra~\cite{Cover--Pombra1989} gives a
multi-letter characterization of the Gaussian feedback capacity, and
Alajaji~\cite{Alajaji1995} characterizes the feedback capacity of
discrete channels with additive noise---feedback does not increase the
capacity of discrete additive channels when there is no input cost
constraint.  Both results can be recast in the form of directed
information (see \cite[Eq.~(52)]{Cover--Pombra1989} and
\cite[Eq.~(17)]{Alajaji1995}).  The notion of directed information in
these contexts, however, has a very limited role as an intermediate
step in the proof of converse coding theorems.  Indeed, the highlight
of Cover--Pombra characterization is the asymptotic equipartition
property of arbitrary nonstationary nonergodic Gaussian
processes~\cite[Section V]{Cover--Pombra1989}; see also
Pinsker~\cite{Pinsker1964}.  (The case of discrete additive channel is
trivial since the optimal input distribution is memoryless and
uniform.)

In a heroic effort~\cite{Tatikonda2000}, Tatikonda attacked the
general nonanticipatory channel with feedback by combining \Verdu--Han
formula for nonfeedback capacity, Massey directed information, and
Shannon strategy for channel side information~\cite{Shannon1958a}.  As
the cost of generality, however, it is extremely difficult to
establish a simple formula like \eqref{eq:multi-fb}.  Furthermore, the
coding theorem in \cite{Tatikonda2000} is not proved in a completely
satisfactory manner.

More recently, Yang, \Kavcic, and
Tatikonda~\cite{Yang--Kavcic--Tatikonda2005} and Chen and
Berger~\cite{Chen--Berger2005} studied special cases of finite-state
channels, based on Tatikonda's framework.  A finite-state
channel~\cite[Section 4.6]{Gallager1968} is described by a conditional
probability distribution
\begin{equation}
\label{eq:fsc}
p(y_n, s_n | x_n s_{n-1}),
\end{equation}
where $s_n$ denotes the channel state at time $n$.  Using a different
approach based on Gallager's proof of the nonfeedback
capacity~\cite[Section 5.9]{Gallager1968}, Permuter, Weissman, and
Goldsmith~\cite{Permuter--Weissman--Goldsmith2006} proved various
coding theorems for finite-state channels with feedback that include
{\it inter alia} the results of \cite{Yang--Kavcic--Tatikonda2005,
Chen--Berger2005} and establish the validity of \eqref{eq:multi-fb}
for indecomposable finite-state channels without intersymbol
interference (i.e., the channel states evolve as an ergodic Markov
chain, independent of the channel input).

As mentioned before, we strive to give a straightforward treatment of
the feedback coding theorem.  Towards this goal, this paper focuses on
stationary nonanticipatory channels of the form
\begin{equation}
\label{eq:channel}
Y_n = g(X_{n-m}, X_{n-m+1}, \ldots, X_n, Z_{n-m}, Z_{n-m+1}, \ldots,
Z_n).
\end{equation}
In words, the channel output $Y_n$ at time $n$ is given as a
time-invariant deterministic function of channel inputs $X_{n-m}^n =
(X_{n-m}, X_{n-m+1}, \ldots, X_n)$ up to past $m$ symbols and channel
noises $Z_{n-m}^n = (Z_{n-m}, Z_{n-m+1}, \ldots, Z_n)$ up to past $m$
symbols.  We assume the noise process $\{Z_n\}_{n=1}^\infty$ is an
arbitrary stationary ergodic process (without any mixing condition)
independent of the message sent over the channel.

The channel model \eqref{eq:channel} is rather simple and physically
motivated.  Yet this channel model is general enough to include many
important feedback communication models such as any additive noise
fading channels with intersymbol interference and indecomposable
finite-state channels without intersymbol interference.%
\footnote{A notable exception is a famous finite-state channel called
the ``trapdoor channel'' introduced by Blackwell~\cite{Blackwell1961}, the
feedback capacity of which is established
in~\cite{Permuter--Cuff--Van-Roy--Weissman2006}.}

The channel~\eqref{eq:channel} has finite input memory in the sense of
Feinstein~\cite{Feinstein1959} and can be viewed as a finite-window
sliding-block coder~\cite[Section 9.4]{Gray1990} of input and noise
processes (cf.\@ primitive channels introduced by Neuhoff and
Shields~\cite{Neuhoff--Shields1979} in which the noise process is
memoryless).  Compared to the general finite-state channel
model~\eqref{eq:fsc} in which the channel has infinite input memory
but the channel noise is memoryless, our channel
model~\eqref{eq:channel} has finite input memory but the noise has
infinite memory; recall that there is no mixing condition on the noise
process $\{Z_n\}_{n=1}^\infty$.  Thus, the finite-state channel model
and the finite sliding-block channel model nicely complement each
other.

\begin{figure}[!t]
\begin{center}
\input{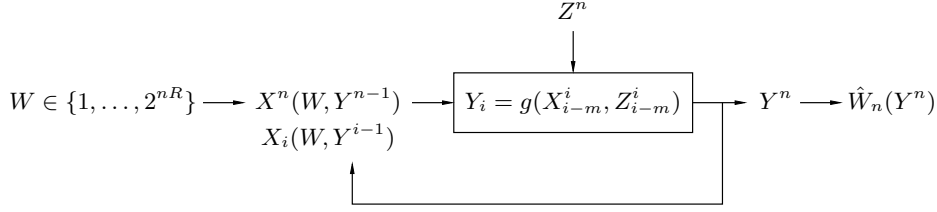}
\end{center}
\caption{Feedback communication channel $Y_i = g(X_{i-m}^i,
Z_{i-m}^i)$.}
\label{fig:channel}
\end{figure}

Our main result is to show that the feedback capacity $C_\textsl{FB}$
of the channel \eqref{eq:channel} is characterized by
\eqref{eq:multi-fb}.  More precisely, we consider a communication
problem depicted in Figure~\ref{fig:channel}.
Here one wishes to
communicated a message index $W \in \{1,2,\ldots, 2^{nR}\}$ over the
channel
\begin{equation}
\label{eq:channel-formal}
Y_i = \left\{
\begin{array}{ll}
\emptyset, & i = 1,\ldots, m,\\
g(X_{i-m}^i, Z_{i-m}^i), & i = m+1, m+2, \ldots,
\end{array}
\right.
\end{equation}
where the time-$i$ channel output $Y_i$ on the output alphabet
$\mathcal{Y}$ is given by a deterministic map $f: \mathcal{X}^m \times
\mathcal{Z}^m \to \mathcal{Y}$ of the current and past $m$ channel
inputs $X_{i-m}^i$ on the input alphabet $\mathcal{X}$ and the current
and past $m$ channel noises $Z_{i-m}^i$ on the noise alphabet
$\mathcal{Z}$.  We assume that the channel noise process
$\{Z_i\}_{i=1}^\infty$ is stationary ergodic and is independent of the
message $W$.  The initial values of $Y_1,\ldots, Y_m$ are set
arbitrarily.  They depend on the unspecified initial condition
$(X_{-m+1}^0, Z_{-m+1}^0)$, the effect of which vanishes from time
$m+1$. Thus the long term behavior of the channel is independent of
$Y_1^m$.

We specify a $(2^{nR}, n)$ feedback code with the encoding maps
\[
X^n(W, Y^{n-1}) = (X_1(W), X_2(W, Y_1), \ldots, X_n(W, Y^{n-1})),
\qquad W = 1,\ldots, 2^{nR},
\]
and the decoding map
\[
\hat{W}_n: \mathcal{Y}^n \to \{1,\ldots, 2^{nR}\}.
\]
The probability of error $P_e^{(n)}$ is defined as
\begin{align*}
P_e^{(n)}  &= \frac{1}{2^{nR}} \sum_{w=1}^{2^{nR}}
\Pr \{ \hat{W}_n(Y^n) \ne w | X^n = X^n(w,Y^{n-1}) \} \\
&= \Pr\{ \hat{W}_n(Y^n) \ne W \},
\end{align*}
where the message $W$ is uniformly distributed over $\{1,\ldots,
2^{nR}\}$ and is independent of $\{Z_i\}_{i=1}^\infty$.  We say that
the rate $R$ is {achievable} if there exists a sequence of $(2^{nR},
n)$ codes with $P_e^{(n)} \to 0$ as $n \to \infty$.  The {feedback
capacity} $C_\textsl{FB}$ is defined as the supremum of all achievable
rates.  The nonfeedback capacity $C$ is defined similarly, with
codewords $X^n(W) = (X_1(W),\ldots, X_n(W))$ restricted to be
a function of the message $W$ only.

We will prove the following result in
Section~\ref{sec:feedback}.
\begin{theorem}
\label{thm:feedback}
The feedback capacity $C_\textsl{FB}$ of the channel
\eqref{eq:channel-formal} is given by
\begin{equation}
C_\textsl{FB} = 
\limp_{n\to\infty} \sup_{p(x^n||y^{n-1})}
\frac{1}{n} I(X^n \to Y^n) \label{eq:feedback}.
\end{equation}
\end{theorem}

Our development has two major ingredients.  First, we revisit the
communication problem over the same channel without feedback in
Section~\ref{sec:nonfeedback} and prove that the nonfeedback capacity
is given by
\[
C = \limp_{n\to\infty} \sup_{p(x^n)} \frac{1}{n}I(X^n;
Y^n).
\]

Roughly speaking, there are three flavors in the literature for the
achievability proof of nonfeedback capacity theorems.  The first one
is Shannon's original argument~\cite{Shannon1948} based on random
codebook generation, asymptotic equipartition property, and joint
typicality decoding, which was made rigorous by
Forney~\cite{Forney1972} and Cover~\cite{Cover1975b}, and now is used
widely in coding theorems for memoryless networks~\cite[Chapter
15]{Cover--Thomas2006}.  This approach, however, does not easily
generalize to channels with memory.  The second flavor is the method
of random coding exponent by Gallager~\cite{Gallager1965}, which was
later applied to finite-state channels~\cite[Section
5.9]{Gallager1968}.  This approach is perhaps the simplest one for the
analysis of general finite-state channels and has been adapted by
Lapidoth and Telatar~\cite{Lapidoth--Telatar1998} for compound
finite-state channels and by Permuter {\it et
al.}~\cite{Permuter--Weissman--Goldsmith2006} for finite-state
channels with feedback.

The third and the least intuitive approach is Feinstein's fundamental
lemma~\cite{Feinstein1954}.  This is the most powerful and general
method of the three, and has been applied extensively in the
literature, say, from Khinchin~\cite{Khinchin1957} to
Gray~\cite{Gray1990} to \Verdu{} and Han~\cite{Verdu--Han1994}.

Our approach is somewhat different from these three usual approaches.
We use the strong typicality (relative frequency) decoding for
$n$-dimensional super letters.  A constructive coding scheme (up to
the level of Shannon's random codebook generation) based on block
ergodic decomposition of Nedoma~\cite{Nedoma1963} is developed, which
uses a long codeword on the $n$-letter super alphabet, constructed as
a concatenation of $n$ shorter codewords.  While each short codeword
and the corresponding output fall into their own ergodic mode, the
long codeword as a whole maintains the ergodic behavior.  To be fair,
codebook construction of this type is far from new in the literature,
and our method is intimately related to the one used by
Gallager~\cite[Section 9.8]{Gallager1968} and Berger~\cite[Section
7.2]{Berger1971} for lossy compression of stationary ergodic sources.
Indeed, when the channel~\eqref{eq:channel} has zero memory ($m = 0$),
then the role of the input for our channel coding scheme is equivalent
to the role of the covering channel for Gallager's source coding
scheme.

Equipped with this coding method for nonfeedback sliding-block coder
channels~\eqref{eq:channel}, the extension to the feedback case is
relatively straightforward.  The basic ingredient for this extension
is the Shannon strategy for channels with causal side information at
the transmitter~\cite{Shannon1958a}.  As a matter of fact, Shannon
himself observed that the major utility of his result is feedback
communication.  Following is the first sentence of
\cite{Shannon1958a}:
\begin{quote}
Channels with feedback from the receiving to the transmitting point
are a special case of a situation in which there is additional
information available at the transmitter which may be used as an aid
in the forward transmission system.
\end{quote}

As observed by Caire and Shamai~\cite[Proposition
1]{Caire--Shamai1999}, the causality has no cost when the transmitter
and the receiver share the same side information---in our case, the
past input (if decoded faithfully) and the past output (received from
feedback)---and the transmission can fully utilize this side
information as if it were known \emph{a priori}.

Intuitively speaking, we can achieve the rate $R_i$ for the $i$th
symbol in the length-$n$ super symbol as
\[
R_i = \max_{p(x^i||y^{i-1})} I(X_i; Y_{i}^n | X^{i-1}, Y^{i-1}),
\qquad i = 1,\ldots, n,
\]
and hence the total achievable rate becomes
\[
R = \max_{p(x^n||y^{n-1})} 
\sum_{i=1}^n I(X_i; Y_{i}^n | X^{i-1}, Y^{i-1})
\]
per $n$ transmissions.  Now a simple algebra shows that this rate is
equal to the maximum directed information as follows:
\begin{align}
\sum_{i=1}^n I(X_i; Y_{i}^n | X^{i-1}, Y^{i-1}) \nonumber
&= \sum_{i=1}^n \sum_{j=i}^n I(X_i; Y_j | X^{i-1}, Y^{j-1})\nonumber\\
&= \sum_{j=1}^n \sum_{i=1}^j I(X_i; Y_j | X^{i-1}, Y^{j-1})\nonumber\\
&= \sum_{j=1}^n I(X^j; Y_j | Y^{j-1})\nonumber\\
&= I(X^n \to Y^n). \label{eq:directed}
\end{align}

The above argument, while intuitively appealing, is not completely
rigorous, however. Therefore, we will take more careful steps, by
first proving the achievability of $\frac{1}{n} I(U^n; Y^n)$ for all
auxiliary random variables $U^n$ and Shannon strategies $X_i(U_i,
X^{i-1}, Y^{i-1}),
\enspace i = 1,\ldots, n,$ and then showing that $I(U^n; Y^n)$ reduces
to $I(X^n \to Y^n)$ via pure algebra.

The next section collects all necessary lemmas that will be used
subsequently in Section~\ref{sec:nonfeedback} for the nonfeedback
coding theorem and in Section~\ref{sec:feedback} for the feedback
coding theorem.

\section{Preliminaries}
\label{sec:lemmas}

Here we review relevant materials from ergodic theory and information
theory in the form of 10 lemmas.  While some of the lemmas are
classical and are presented in order to make the paper self-contained,
the other lemmas are crucial to our main discussion in subsequent
sections and may contain original observations.  Throughout this
section, $\mathbf{Z} = \{Z_i\}_{i=1}^\infty$ denotes a generic
stochastic process on a finite alphabet $\mathcal{Z}$ with associated
probability measure $P$ defined on Borel sets under the usual topology
on $\mathcal{Z}^{\infty}$.

\subsection{Ergodicity}
Given a \emph{stationary} process $\mathbf{Z} = \{Z_i\}_{i=1}^\infty$,
let $T: \mathcal{Z}^\infty \to \mathcal{Z}^\infty$ be the
associated measure preserving shift transformation.  Intuitively, $T$
maps the infinite sequence $(z_1, z_2, z_3, \ldots)$ to $(z_2, z_3,
z_4, \ldots)$.  We say the transformation $T$ (or the process
$\mathbf{Z}$ itself) is \emph{ergodic} if every measurable set $A$
with $TA = A$ satisfies either $P(A) = 0$ or $P(A) = 1$.

The following characterization of ergodicity is well known; see, for
example, Petersen~\cite[Exercise 2.4.4]{Petersen1983} or
Wolfowitz~\cite[Lemma 10.3.1]{Wolfowitz1978}.
\begin{lemma}
\label{lemma:ergodicity}
Suppose $\{Z_i\}_{i=1}^\infty$ be a stationary process and let $T$
denote the associated measure preserving shift transformation.
Then, $\{Z_i\}$ is ergodic if and only if
\[
\lim_{n\to\infty} \frac{1}{n} \sum_{i=0}^{n-1} P(T^{-i} A \cap B) =
P(A)\cdot P(B)
\qquad\text{ for all measurable $A$ and $B$.}
\]
\end{lemma}

When $\mathbf{X} = \{X_i\}_{i=1}^\infty$ and $\mathbf{Z} =
\{Z_i\}_{i=1}^\infty$ are independent stationary ergodic processes,
they are not necessarily jointly ergodic.  For example, if we take
\[
\mathbf{X} = 
\left\{
\begin{matrix}
01010101\ldots, &\quad{\text{with probability }} 1/2,\\
10101010\ldots, &\quad{\text{with probability }} 1/2,\\
\end{matrix}
\right.
\]
and $\mathbf{Z}$ is independent and identically distributed as
$\mathbf{X}$, then it is easy to verify that $\{Y_i = X_i + Z_i
\pmod{2}\}_{i=1}^\infty$ is \emph{not} ergodic.  However, if one of
the processes is mixing reasonably fast, then they are jointly
ergodic.  The following result states a sufficient condition for joint
ergodicity.
\begin{lemma}
\label{lemma:product}
If\, $\mathbf{X}$ is independent and identically distributed (i.i.d.),
and\, $\mathbf{Z}$ is stationary ergodic, independent of\,
$\mathbf{X}$, then the pair $(\mathbf{X}, \mathbf{Z}) = \{(X_i,
Z_i)\}_{i=1}^\infty$ is jointly stationary ergodic.
\end{lemma}
A stronger result is true, which assumes $\mathbf{X}$ to be weakly
mixing only.  The proof is an easy consequence of
Lemma~\ref{lemma:ergodicity}; for details refer to
Brown~\cite[Proposition 1.6]{Brown1976} or Wolfowitz~\cite[Theorem
10.3.1]{Wolfowitz1978}.

We will later need to construct \emph{super-letter} processes for our
coding theorems.  The next lemma due to Gallager~\cite[Lemma
9.8.2]{Gallager1968} deals with the ergodic decomposition of the
$n$-letter super process that is built from a single-letter stationary
ergodic one; see also Nedoma~\cite{Nedoma1963} and
Berger~\cite[Section 7.2]{Berger1971}.
\begin{lemma}
\label{lemma:super}
Suppose $\mathbf{Z} = \{Z_i\}_{i=1}^\infty$ be stationary ergodic on
$\mathcal{Z}$, and let $T$ be the associated shift transformation.
Define the $n$th-order super process $\mathbf{Z}^{(n)} =
\{Z_i^{(n)}\}_{i=1}^\infty$ on $\mathcal{Z}^n$ as
\[
Z_i^{(n)} = (Z_{n(i-1)+1}, Z_{n(i-1)+2},\ldots, Z_{ni}),\qquad i =
1,2,\ldots.
\]
Then, the super process $\mathbf{Z}^{(n)}$ has $n'$ ergodic modes,
each with probability $1/n'$ and disjoint up to measure zero, where
$n'$ divides $n$.  Furthermore, in the space $\mathcal{Z}^\infty$
of the original process $\mathbf{Z}$, the sets $S_1, S_2, \ldots,
S_{n'}$ corresponding to these ergodic modes can be related by $T(S_i)
= S_{i+1},\enspace 1 \le i \le n-1,$ and $T(S_n) = S_1$.
\end{lemma}

We will use the notation $P(\cdot|S_k),\enspace k = 1,\ldots, n',$ for
the probability measure under each ergodic mode.

\subsection{Strong Typicality}
We use the strong typicality~\cite[Section 10.6]{Cover--Thomas2006} as
the basic method of decoding.  Here we review a few basic properties
of strongly typical sequences.

First definitions.  Let $N(a|x^n)$ denote the number of occurrences of
the symbol $a$ in the sequence $x^n$.  We say a sequence $x^n \in
\mathcal{X}^n$ is \emph{$\epsilon$-strongly typical} (or typical in
short) with respect to a distribution $P(x)$ on $\mathcal{X}$ if
\[
\left| \frac{1}{n} N(a|x^n) - P(a)\right | < \frac{\epsilon}{|\mathcal{X}|}
\]
for all $a \in \mathcal{X}$ with $P(a) > 0$, and $N(a|x^n) = 0$ for
all $a \in \mathcal{X}$ with $P(a) = 0$.  Consistent with this
definition, we say a pair of sequences $(x^n, y^n)$ are \emph{jointly
$\epsilon$-strongly typical} (or jointly typical in short) with
respect to a distribution $P(x,y)$ on $\mathcal{X}\times\mathcal{Y}$
if
\[
\left| \frac{1}{n} N(a,b|x^n,y^n) - P(a,b)\right | <
\frac{\epsilon}{|\mathcal{X}||\mathcal{Y}|}
\]
for all $(a,b) \in \mathcal{X}\times\mathcal{Y}$ with $P(a,b) > 0$,
and $N(a,b|x^n) = 0$ for all $(a,b) \in \mathcal{X}\times\mathcal{Y}$
with $P(a,b) = 0$.

The set of strongly typical sequences $x^n \in \mathcal{X}^n$ with
respect to $X \sim P(x)$ is denoted $A_\epsilon^{*(n)}(X)$.  We
similarly define a joint typical set $A_\epsilon^{*(n)}(X,Y)$
for $(X,Y) \sim P(x,y)$.

The following statement is a trivial consequence of the definition of
typical sequences.
\begin{lemma}
\label{lemma:function}
Suppose $X \sim P(X)$.  If $x^n \in A_\epsilon^{*(n)}(X)$ and $y^n =
f(x^n) := (f(x_1), f(x_2), \ldots, f(x_n))$, then $y^n \in
A_\delta^{*(n)}(f(X))$ with $\delta = \epsilon\cdot(|\mathcal{X}|-1)$.
\end{lemma}
As a special case, if $(x^n, y^n)$ is $\epsilon$-strongly typical with
respect to a joint distribution $P(x,y)$, then $x^n$ is
$\epsilon$-strongly typical with respect to the marginal $P(x) =
\sum_y P(x,y)$.

Our discussion on the typical sequences so far has not given a
specific context on how they are generated.  Now we connect the notion
of strong typicality with ergodic processes.  First, from Birkhoff's
ergodic theorem~\cite[Theorem 2.2.3]{Petersen1983} and the definition
of ergodicity, the following lemma is immediate.
\begin{lemma}
Let $\mathbf{Z} = \{Z_i\}_{i=1}^\infty$ be stationary ergodic with
$Z_1 \sim P(z)$.  Then
\[
\Pr(Z^n \in A_\epsilon^{*(n)}(Z_1)) \to 1
\qquad\text{ as $n \to \infty$.}
\]
\end{lemma}

\begin{figure}[!t]
\begin{center}
\input{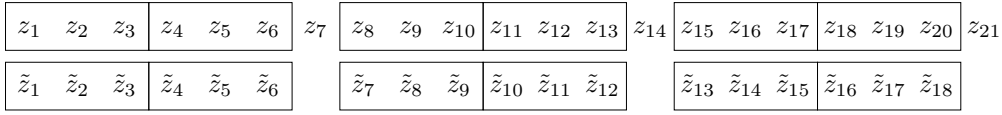}
\end{center}
\caption{Construction of $\tilde{Z}^{Ln^2}$ from $Z^{Ln^2+n}$: $n = 3,
L = 2$.}
\label{fig:const1}
\end{figure}

As we mentioned in the previous subsection, the $n$th order super
process $\mathbf{Z}^{(n)} = \{Z_i^{(n)}\}_{i=1}^\infty$ defined as
\[
Z_i^{(n)} = Z_{(n-1)i+1}^{ni}, \qquad i=1,2,\ldots,
\]
is not necessarily ergodic, but is a mixture of disjoint ergodic
modes.  Thus, the super process $\mathbf{Z}^{(n)}$ is not necessarily
typical with respect to $P(z^n)$ on the $n$-letter alphabet
$\mathcal{Z}^n$.  The following construction by
Gallager~\cite[pp.~498--499]{Gallager1968}, however, gives a typical
sequence in the $n$-letter super alphabet by shifting through each
ergodic phase.
\begin{lemma}
\label{lemma:gallager}
Given positive integers $n, L$ and a stationary ergodic process
$\mathbf{Z} = \{Z_i\}_{i=1}^\infty$, construct $\tilde{\mathbf{Z}} =
\{\tilde{Z}_i\}_{i=1}^{Ln^2}$ as follows (See Figure~\ref{fig:const1}):
\[
\tilde{Z}_i = 
\left \{
\begin{array}{ll}
Z_i, & \quad i = 1,\ldots, Ln,\\
Z_{i+1}, &\quad i = Ln+1, \ldots, 2Ln,\\
&\vdots\\
Z_{i+n-1}, &\quad i = Ln(n-1) + 1, \ldots, Ln^2.\\
\end{array}
\right.
\]
In other words, $\{\tilde{Z}_i\}_{i=1}^{Ln^2}$ is a verbatim copy of
$\{Z_i\}_{i=1}^{Ln^2+n}$ with every $(Ln+1)$st position skipped.  Let
$\tilde{\mathbf{Z}}^{(n)} = (\tilde{Z}_1^n, \tilde{Z}_{n+1}^{2n},
\ldots, \tilde{Z}_{Ln^2 - n + 1}^{Ln^2})$ be the associated $n$th
order super process of length $Ln$.  Then,
\[
\Pr(\tilde{\mathbf{Z}}^{(n)} \in A_\epsilon^{*(Ln)}(Z^n)) \to 1
\qquad\text{ as $L \to \infty$}.
\]
\end{lemma}

\begin{proof}
From Lemma~\ref{lemma:super} and the given construction of skipping
one position after every $Ln$ symbols, each of $n$ sequences
\begin{align*}
(\tilde{Z}_1^n, \ldots, \tilde{Z}_{Ln-n+1}^{Ln}) 
&= (Z_1^n, \ldots, Z_{Ln-n+1}^{Ln})\\
(\tilde{Z}_{Ln+1}^{Ln+n}, \ldots, \tilde{Z}_{2Ln-n+1}^{2Ln}) 
&= (Z_{Ln+2}^{L(n+1)+1}, \ldots, Z_{2Ln-n+2}^{2Ln+1})\\
&\,\,\;\vdots\\
(\tilde{Z}_{Ln(n-1)+1}^{Ln(n-1)+n}, \ldots, \tilde{Z}_{Ln^2-n+1}^{Ln^2}) 
&= (Z_{Ln(n-1)+n}^{Ln(n-1)+2n-1}, \ldots, Z_{Ln^2}^{Ln^2+n-1})
\end{align*}
falls in one of ergodic modes $(S_1,\ldots, S_{n'})$ with $n/n'$
sequences for each mode.  Now for each sequence with corresponding
ergodic mode $S_k$, the relative frequencies of all super symbols $a^n
\in \mathcal{Z}^n$ converge to the corresponding distribution
$P(a^n|S_k)$ as $L \to \infty$.  But each ergodic mode is visited
evenly, each by $n/n'$ sequences.  Therefore, the relative frequencies
of all $a^n \in \mathcal{Z}^n$ in the entire sequence
$\tilde{Z}_1^{Ln^2}$ converge to
\[
\frac{1}{n'} \sum_{k=1}^{n'} P(a^n|S_k) = P(a^n)
\]
as $L \to \infty$.
\end{proof}

Combining Lemma~\ref{lemma:product} with the proof of
Lemma~\ref{lemma:gallager}, we have the following result.
\begin{lemma}
\label{lemma:joint-super}
Under the condition of Lemma~\ref{lemma:gallager}, let further
$\mathbf{X} = \{X_i\}_{i=1}^\infty$ be blockwise \iid $\sim
P(x^n)$, that is, $X_i^{(n)} = X_{(n-1)i+1}^{ni}, \enspace
i=1,2,\ldots, \iid \sim P(x^n)$, independent of\, $\mathbf{Z}$.  Then,
\[
\Pr((\mathbf{X}^{(n)}, \tilde{\mathbf{Z}}^{(n)}) \in
A_\epsilon^{*(Ln)}(X^n,Z^n) ) \to 1\qquad
\text{ as $L \to \infty$}.
\]
\end{lemma}

Finally we recall the key result linking the typicality with mutual
information~\cite[Lemma 10.6.2]{Cover--Thomas2006}.
\begin{lemma}
\label{lemma:joint-typicality}
Suppose $(X,Y) \sim P(x,y)$ and let $X_1,X_2, \ldots, X_n$ be \iid
$\sim P(x)$.  For $y^n \in A_\epsilon^{*(n)}(Y)$, the probability that
$(X^n, y^n) \in A_\epsilon^{*(n)}(X,Y)$ is upper bounded by
\[
\Pr((X^n, y^n) \in A_\epsilon^{*(n)}(X, Y)) \le 2^{-n(I(X;Y) - \delta)}
\]
where $\delta \to 0$ as $\eps \to 0$.
\end{lemma} 

\subsection{Channels with Side Information}
We prove the following identity in a purely algebraic manner, then
find its meaning in information theory.  Here we assume every alphabet
is finite.
\begin{lemma}
\label{lemma:state1}
Suppose $S \sim p(s)$.  For a given conditional distribution
$p(y|x,s)$ on the product space $\mathcal{X}\times \mathcal{S} \times
\mathcal{Y}$, we have
\begin{equation}
\label{eq:state1}
\max_{p(u),x = f(u,s)} I(U; Y, S) = \max_{p(x|s)} I(X;Y|S),
\end{equation}
where the maximum on the left hand side is taken over all conditional
distributions of the form $p(u,x|s) = p(u) p(x|u,s)$ with
deterministic $p(x|u,s)$ (that is, $p(x|u,s) = 0$ or $1$), and the
auxiliary random variable $U$ has cardinality bounded by
$|\mathcal{U}| \le (|\mathcal{X}|-1)|\mathcal{S}|$.
\end{lemma}
\begin{proof}
For any joint distribution of the form $p(u,x,s,y) = p(u)p(s)p(x|u,s)
p(y|x,s)$ with deterministic $p(x|u,s)$, we have the following Markov
chains: $U \to (X,S) \to Y$ and $X \to (U,S) \to Y$.  Combined with
the independence of $U$ and $S$, these Markov relationships imply that
\begin{align}
\max_{p(u),x = f(u,s)} I(U; Y, S) \label{eq:u-ys}
&=\max_{p(u),x = f(u,s)} I(U; Y| S) \\
&=\max_{p(u), x = f(u,s)} I(X;Y|S).\notag
\end{align}
But it can be easily verified that any conditional distribution
$p(x|s)$ can be represented as 
\[
p(x|s) = \sum_u p(u) p(x|u,s)
\]
for appropriately chosen $p(u)$ and \emph{deterministic} $p(x|u,s)$
with cardinality of $U$ upper bounded by $|\mathcal{U}| \le 
(|\mathcal{X}|-1)|\mathcal{S}|$.  Therefore, we have
\[
\max_{p(u), x = f(u,s)} I(X;Y|S) = \max_{p(x|s)} I(X; Y|S),
\]
which proves the desired result.
\end{proof}

It is well known that the capacity of a memoryless state-dependent
channel $p(y|x,s)$ is given as
\[
C = \max_{p(x|s)} I(X;Y | S),
\]
if the state information is known at both the encoder and decoder
prior to the actual communication.  What will happen if the
transmitter learns the state information on the fly, so that only the
past and present state realization can be utilized for communication?

\begin{figure}[!t]
\begin{center}
\input{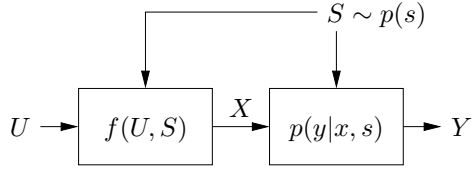}
\end{center}
\caption{Shannon strategy for coding with side information.}
\label{fig:shannon}
\end{figure}

Shannon~\cite{Shannon1958a} considered the communication over a
memoryless state-dependent channel $p(y|x,s)$ with state information
available \emph{only} at the transmitter on the fly, and showed that
the capacity is given by
\begin{equation}
\label{eq:shannon}
C = \max_{p(u),x=f(u,s)} I(U;Y),
\end{equation}
where the cardinality of $U$ is bounded as $|\mathcal{U}| \le
|\mathcal{X}|^{|\mathcal{S}|}$, counting for all functions $f:
\mathcal{S} \to \mathcal{X}$.  This capacity is achieved by attaching
a physical device $X = f(U,S)$ in front of the actual channel as
depicted in Figure~\ref{fig:shannon}, which maps the channel state $S$
to the channel input $X$ according to the function (index) $U$.
Now treating $U$ as the input to the newly generated channel
\[
p(y|u) = \sum_{x,s} p(s) p(x|u,s) p(y|x,s)
\]
and coding as in the case of usual memoryless channels, we can easily
achieve $I(U;Y)$.  This method, surprisingly simple yet optimal, is
sometimes called the Shannon strategy.

Now when the decoder also knows the channel state $S$, it is
equivalent for the decoder to receive the augmented channel output
${Y}' = (Y,S)$.  Thus, the capacity of the same channel $p(y|x,s)$
with the state information \emph{causally} known at both the encoder
and decoder%
\footnote{For the usual block coding, the decoder causality is
irrelevant. The message is decoded only after the entire block is
received.}
follows
from \eqref{eq:shannon} as 
\[
C = \max_{p(u),x=f(u,s)} I(U;Y,S).
\]
Therefore, Lemma~\ref{lemma:state1} states that when the same side
information is available at the receiver, the causal encoder with the
best Shannon strategy performs no worse than the noncausal encoder who
can preselect the entire codeword compatible with the whole state
sequence.

For the last lemma needed for main results, we recall the notation of
causally conditioned distributions
\begin{align}
\label{eq:causal1}
p(x^n||y^{n-1}) &= \prod_{i=1}^n p(x_i|x^{i-1}, y^{i-1})
\intertext{and}
\label{eq:causal2}
p(y^n||x^n) &= \prod_{i=1}^n p(y_i|x^{i}, y^{i-1}).
\end{align}
(The notation \eqref{eq:causal1} and \eqref{eq:causal2} can be unified
if we define 
\[
p(a^n||b^m) = \left\{
\begin{array}{ll}
\prod_{i=1}^n p(a_i|b^{i}, a^{i-1}),&\qquad n = m,\\
p(a^n||\emptyset^{n-m} b^m), &\qquad n > m,\\
p(\emptyset^{m-n} a^n || b^m), &\qquad n < m.)
\end{array}\right.
\]
By chain rule, we have
\[
p(x^n||y^{n-1}) p(y^n||x^n) = p(x^n, y^n) = p(x^n) p(y^n|x^n)
\]
for any joint distribution $p(x^n, y^n)$.  Thus, given a causally
conditioned distribution (or a channel) $p(y^n||x^n)$, the causally
conditioned distribution (or the input) $p(x^n||y^{n-1})$ completely
specifies the joint distribution $p(x^n, y^n)$.

As a corollary of Lemma~\ref{lemma:state1}, we have the following
result.
\begin{lemma}
\label{lemma:state2}
Suppose a causally conditioned distribution $p(y^n||x^n)$ is given.
Then we have
\begin{equation}
\label{eq:state2}
\max_{p(u^n), x_i = f(u_i, x^{i-1}, y^{i-1})} I(U^n; Y^n) =
\max_{p(x^n||y^{n-1})} I(X^n \to Y^n),
\end{equation}
where the maximum on the left hand side is taken over all 
joint distributions of the form
\begin{align}
p(u^n, x^n, y^n) &= \prod_{i=1}^n
\bigl(
p(u_i) p(x_i | u_i, x^{i-1}, y^{i-1}) p(y_i|x^{i}, y^{i-1})
\bigr) \notag\\
&= \Bigl(\prod_{i=1}^n
p(u_i) p(x_i | u_i, x^{i-1}, y^{i-1}) 
\Bigr)
p(y^n||x^n)
\label{eq:dist}
\end{align}
with deterministic $p(x_i|u_i, x^{i-1}, y^{i-1})$, $i=1,\ldots,n$, and
the auxiliary random variables $U_i$ has the cardinality bounded by
$|\mathcal{U}_i| \le |\mathcal{X}|^{i}|\mathcal{Y}|^{i-1}$.
\end{lemma}
\begin{proof}
Let $q(u^n, x^n, y^n)$ be any joint distribution of the form
\eqref{eq:dist} such that $q(x_i|u_i,x^{i-1}, y^{i-1})$, $i=1,\ldots,
n$ are deterministic and that $q(y^n||x^n) = p(y^n||x^n)$ (i.e., the
joint distribution $q(u^n,x^n,y^n)$ is consistent with the given
causally conditioned distribution $p(y^n||x^n)$).  For $(U^n, X^n,
Y^n) \sim q(u^n, x^n, y^n)$, it is easy to verify that $U_i^n$ is
independent of $(U^{i-1}, X^{i-1}, Y^{i-1})$, which implies that
$U^{i-1} \to (X^{i-1}, Y^{i-1}) \to Y_i^n$ forms a Markov chain.  On
the other hand, $X^{i-1}$ is a deterministic function of $(U^{i-1},
Y^{i-1})$ and thus $X^{i-1} \to (U^{i-1}, Y^{i-1}) \to Y_i^n$ also
forms a Markov chain.  Similarly, we have the Markovity for $U^i \to
(X^i, Y^{i-1}) \to Y_{i}^n$ and $X^i \to (U^i, Y^{i-1}) \to Y_{i}^n$.
Therefore, we have
\begin{align}
I(U_i; Y^n | U^{i-1}) 
&= I(U_i; Y_i^n | Y^{i-1}, U^{i-1}) \label{eq:first-eq}\\
&= H(Y_i^n | Y^{i-1}, U^{i-1}) - H(Y_i^n | Y^{i-1}, U^i) \notag\\
&= H(Y_i^n | Y^{i-1}, X^{i-1}) - H(Y_i^n | Y^{i-1}, X^i) \label{eq:third-eq}\\
&= I(X_i;  Y_{i}^n | X^{i-1}, Y^{i-1}), \notag
\end{align}
where \eqref{eq:first-eq} follows from the independence of $U_i$ and
$(U^{i-1}, Y^{i-1})$, and \eqref{eq:third-eq} follows from Markov
relationships observed above.  Now from the alternative expansion of
the directed information shown in~\eqref{eq:directed}, we have
\[
\max_{q} I(U^n; Y^n) =
\max_{q} I(X^n \to Y^n).
\]
Finally, by using distributions of the form
\[
p(x_i|x^{i-1}, y^{i-1})
= \sum_{u_i} p(u_i) p(x_i|u_i, x^{i-1}, y^{i-1}),\qquad i = 1,\ldots, n
\]
with appropriately chosen $p(u_i)$ and deterministic
$p(x_i|u_i,x^{i-1}, y^{i-1})$, we can represent any causally conditioned
distribution 
\[
p(x^n||y^{n-1})
= \prod_{i=1}^n
p(x_i|x^{i-1}, y^{i-1})
= \sum_{u^n} \prod_{i=1}^n \bigl( p(u_i) p(x_i|u_i, x^{i-1}, y^{i-1})
\bigr),
\]
which implies that
\[
\max_q I(X^n \to Y^n) = \max_{p(x^n||y^{n-1})} I(X^n \to Y^n)
\]
and completes the proof.
\end{proof}

\section{Nonfeedback Coding Theorem Revisited}
\label{sec:nonfeedback}

This section is devoted to the proof of the following result.

\begin{theorem}
The nonfeedback capacity $C$ of the stationary channel
\begin{equation}
\label{eq:thm-channel}
Y_i = \left\{
\begin{array}{ll}
\emptyset, & i = 1,\ldots, m,\\
g(X_{i-m}^i, Z_{i-m}^i), & i = m+1, m+2, \ldots,
\end{array}
\right.
\end{equation}
with the input $X_i$ and the stationary ergodic noise process
$\{Z_i\}_{i=1}^\infty$ depicted in Figure~\ref{fig:channel} is given by
\begin{align}
C &= \lim_{n\to\infty} C_n \notag\\
&= \limp_{n\to\infty} \sup_{p(x^n)} 
\frac{1}{n} I(X^n; Y^n). \label{eq:nonfeedback}
\end{align}
\end{theorem}

Revisiting and proving the nonfeedback coding theorem is rewarding for
two reasons.  First, our proof is somewhat different from the usual
techniques and hence is interesting on its own.  (See
Section~\ref{sec:intro} for the discussion on conventional
achievability proofs of nonfeedback capacity theorems.)  Second, our
exercise here will lead to a straightforward proof of the feedback
coding theorem in the next section.

\begin{proof}
We first note that the capacity expression \eqref{eq:nonfeedback} is
well-defined because $n C_n$ is superadditive (i.e., $m C_m + n C_n
\le (m+n) C_{m+n}$), which implies that the limit exists and
\[
\lim_{n\to\infty} C_n = \sup_{n \ge 1} C_n.
\]

The converse follows immediately from Fano's inequality~\cite[Lemma
  7.9.1]{Cover--Thomas2006}.  For any sequence of $(2^{nR}, n)$ codes
$(X^n(W), \hat{W}(Y^n))$ with the message $W$ drawn uniformly over
$\{1,\ldots,2^{nR}\}$, if
\[
P_e^{(n)} = \Pr(W \ne \hat{W}) \to 0,
\]
then we must have
\begin{align*}
nR &\le I(W; Y^n) + n\eps_n \\
&\le I(X^n; Y^n) + n\eps_n
\end{align*}
where $\eps_n \to 0$ as $n \to \infty$.

For the achievability, it suffices to show that there exists a
sequence of codes that achieves $C_n$ for each $n > m$.  (Recall $C_1 =
\cdots C_m = 0$.)  Without loss of generality, we assume that the
alphabets $\mathcal{X}, \mathcal{Y}, \mathcal{Z}$ are finite.
Otherwise, we can partition the space for each $n$ and $\eps >0$ 
such that
\[
\max_{p([x]^n)} \frac{1}{n} I([X]^n; [Y]^n) \ge C_n -
\eps,
\]
and prove the achievability on this partitioned space.

{\it Codebook generation.}  Fix $n > m$ and let $p^*(x^n)$ denote the
input distribution that achieves $C_n$.  For each $L = 1,2,\ldots$,
let $k = k(L,n) = Ln^2 + n$.  We generate a sequence%
\footnote{This gives only a \emph{subsequence} of $(2^{kR}, k)$ codes.
But we can easily interpolate to $Ln^2 + n < k < (L+1)n^2 + n$ without
any rate loss, since $(Ln^2+n)/((L+1)n^2+n) \to 1$ as $L \to \infty$.}
of $(2^{kR}, k)$ codes $X^k(w)$ as depicted in
Figure~\ref{fig:const4}.
\begin{figure}[!t]
\begin{center}
\input{const4.pstex_t}
\end{center}
\caption{Input, noise, and output sequences: $n = 3, L = 2, m = 1$.}
\label{fig:const4}
\end{figure}

For each $w \in \{1,2,\ldots,2^{kR}\}$, generate a codeword
$\tilde{\mathbf{X}}^{(n)}(w) = \tilde{X}^{Ln^2}(w)$ of length $Ln$ on
the $n$-letter super alphabet $\mathcal{X}^n$ independently according
to
\[
p(\tilde{x}^{Ln^2}) = \prod_{i=1}^{Ln} p^*(x_{(n-1)i + 1}^{ni}).
\]
We exhibit the $2^{kR}$ codewords as the rows of a matrix:
\[
\mathcal{C} = \left[
\begin{matrix}
\tilde{X}_1^n(1) & \tilde{X}_{n+1}^{2n}(1) & \cdots &
\tilde{X}_{Ln(n-1)+1}^{Ln^2}(1) \\
\vdots & \vdots & \ddots & \vdots \\
\tilde{X}_1^n(2^{kR}) & \tilde{X}_{n+1}^{2n}(2^{kR}) & \cdots &
\tilde{X}_{Ln(n-1)+1}^{Ln^2}(2^{kR})
\end{matrix}
\right].
\]
Each entry in this matrix is generated \iid according to $p^*(x^n)$.

Using the construction as in Lemma~\ref{lemma:gallager} (see
Figure~\ref{fig:const4}), the \emph{actual} codewords ${\mathbf{X}}(w)
= X^k(w),\enspace w = 1,2,\ldots,2^{nR},$ which will be transmitted
over the channel, are generated from $\tilde{\mathbf{X}}^{(n)}(w) =
\tilde{X}^{Ln^2}(w)$ as follows:
\[
{X}_{(i-1)Ln+i}^{iLn+i} 
= \bigl(\tilde{X}_{(i-1)Ln + 1}^{iLn},\; \emptyset\bigr),
\qquad i = 1, 2, \ldots, n.
\]
In other words, $X^k$ is a verbatim copy of $\tilde{X}^{Ln^2}$ with
fixed symbol $\emptyset$ separating the subsequences of length $Ln$.

{\it Encoding.} If $W = w$, the transmitter sends the codeword
$\mathbf{X}(w) = X^k(w)$ over the channel.

{\it Decoding.} Upon receiving the sequence $\mathbf{Y} = Y^k$, the
receiver forms the sequence $\tilde{\mathbf{Y}}^{(n)} =
\tilde{Y}^{Ln^2}$ of length $Ln$ in the $n$-letter super alphabet
$\mathcal{Y}^{n}$, as depicted in Figure~\ref{fig:const4}:
\[
\tilde{Y}_{(i-1)n+1}^{in}
=\left\{
\begin{array}{ll}
(\emptyset, Y_{(i-1)n+m+1}^{in}), &\quad i = 1,\ldots, L,\\
(\emptyset, Y_{(i-1)n+m+2}^{in+1}), &\quad i = L+1,\ldots, 2L,\\
&\vdots \\
(\emptyset, Y_{(i-1)+m+n-1}^{in+n}), & \quad i = L(n-1) + 1, \ldots, Ln.
\end{array}
\right.
\]

Now we consider $\tilde{\mathbf{X}}^{(n)} = \tilde{X}^{Ln^2}$ and
$\tilde{\mathbf{Y}}^{(n)} = \tilde{Y}^{Ln^2}$ as sequences of length
$Ln$ on the super alphabet $\mathcal{X}^n \times \mathcal{Y}^{n}$.
The receiver declares that the message $\hat{W}$ was sent if there is
a unique $\hat{W}$ such that
\[
(\tilde{\mathbf{X}}^{(n)}(\hat{W}), \tilde{\mathbf{Y}}^{(n)}) \in
A_\eps^{*(Ln)}(X^n,Y^n),
\]
that is, $(\tilde{\mathbf{X}}^{(n)}(\hat{W}),
\tilde{\mathbf{Y}}^{(n)})$ is jointly typical with respect to the
joint distribution $p(x^n, y^n)$ specified by $p^*(x^n) p(z^n)$
and the definition of the channel \eqref{eq:thm-channel}.  Otherwise,
an error is declared.

{\it Analysis of the probability of error.}  Without loss of
generality, we assume $W=1$ was sent.  We define the following events:
\[
E_i = \{(\tilde{\mathbf{X}}^{(n)}(1), \tilde{\mathbf{Y}}^{(n)})
\in A_\eps^{*(Ln)}(X^n, Y^n)\},\qquad
i \in \{1,2,\ldots, 2^{kR} \},
\]
where $E_i$ is the event that the $i$th codeword and
$\tilde{\mathbf{Y}}^{(n)})$ are jointly typical.  By Bonferonni's
inequality, we have
\begin{align*}
\Pr(\hat{W} \ne W) &= \Pr(\hat{W} \ne W | W = 1)\\
&= \Pr(E_1^c \cup E_2 \cup E_3 \cup \cdots \cup E_{2^{kR}}) \\
&\le \Pr(E_1^c) + \sum_{i=2}^{2^{kR}} \Pr(E_i).
\end{align*}

In order to bound $\Pr(E_1^c)$, we define $\tilde{\mathbf{Z}}^{(n)}$
as the $n$th order super process of length $Ln$ on the super alphabet
$\mathcal{Z}^n$ constructed from the noise process
$\{Z_i\}_{i=1}^\infty$ as in Lemma~\ref{lemma:gallager}.  (See
Figure~\ref{fig:const4}.)  Since $\tilde{\mathbf{X}}^{(n)}(1)$ is
blockwise \iid $\sim p^*(x^n)$ and independent of $\mathbf{Z}$, we
have from Lemma~\ref{lemma:joint-super}
\[
\Pr((\tilde{\mathbf{X}}^{(n)}(1), \tilde{\mathbf{Z}}^{(n)}) \in
A_\epsilon^{*(Ln)}(X^n,Z^n) ) \to 1\qquad \text{ as } L \to \infty.
\]
Furthermore, $\tilde{\mathbf{Y}}^{(n)}$ is the
blockwise function of $(\tilde{\mathbf{X}}^{(n)}(1),
\tilde{\mathbf{Z}}^{(n)})$, that is,
\[
\tilde{Y}_{(i-1)n+1}^{in} = f(\tilde{X}_{(i-1)n+1}^{in}(1),
\tilde{Z}_{(i-1)n+1}^{in})
\]
with the time-invariant function $f$ induced by the channel function
$g$ in~\eqref{eq:thm-channel}.  Thus by Lemma~\ref{lemma:function},
\[
\Pr((\tilde{\mathbf{X}}^{(n)}(1), \tilde{\mathbf{Y}}^{(n)}) \in
A_\epsilon^{*(Ln)}(X^n,Y^n) ) \to 1\qquad\text{ as } L \to\infty,
\]
and
\begin{equation*}
\Pr(E_1^c) \le \eps \qquad\text{for $L$
sufficiently large.}
\end{equation*}

On the other hand, recall that the typicality of
$(\tilde{\mathbf{X}}^{(n)}(i), \tilde{\mathbf{Y}}^{(n)})$ implies
the typicality of $\tilde{\mathbf{Y}}^{(n)}$
(Lemma~\ref{lemma:function}).  Hence, by
Lemma~\ref{lemma:joint-typicality} we have for each $i \ne 1$
\begin{align*}
\Pr(E_i)
&= \Pr((\tilde{\mathbf{X}}^{(n)}(i), \tilde{\mathbf{Y}}^{(n)})
\in A_\eps^{*(Ln)})\notag\\
&= \sum_{\tilde{\mathbf{y}}^{(n)} \in A_\eps^{*(Ln)}}
  \Pr((\tilde{\mathbf{X}}^{(n)}(i), \tilde{\mathbf{y}}^{(n)})
&\le 2^{-Ln (I(X^n; Y_{m+1}^n)-\delta)},
\end{align*}
where $\delta \to 0$ as $\eps \to 0$.
Consequently,
\begin{align*}
\Pr(\hat{W} \ne W) &\le \Pr(E_1^c) + \sum_{i=2}^{2^{kR}} \Pr(E_i)\\
&\le \eps + 2^{kR} 2^{-Ln (I(X^n; Y_{m+1}^n)-\delta)} \\
&\le 2 \eps
\end{align*}
if $L$ is sufficiently large and 
\[
kR < Ln (I(X^n; Y_{m+1}^n) - \delta),
\]
or equivalently,
\[
R < \frac{Ln^2+n}{Ln} (I(X^n; Y_{m+1}^n) - \delta).
\]
Since $\eps$ can be made arbitrarily small and
$(Ln^2+n)/(Ln) \to 1/n$ as $L\to \infty$, we have a sequence of
$(2^{kR}, k)$ codes that achieves 
\[
R < \frac{1}{n} I(X^n; Y_{m+1}^n) = \frac{1}{n} I(X^n; Y^n) = C_n.
\]
\end{proof}

\section{Proof of Theorem~\ref{thm:feedback}}
\label{sec:feedback}
Recall our channel model:
\begin{equation}
\label{eq:thm-channel-fb}
Y_i = \left\{
\begin{array}{ll}
\emptyset, & i = 1,\ldots, m,\\
g(X_{i-m}^i, Z_{i-m}^i), & i = m+1, m+2, \ldots,
\end{array}
\right.
\end{equation}
with the input $X_i$ and the stationary ergodic noise process
$\{Z_i\}_{i=1}^\infty$ depicted in Figure~\ref{fig:channel}.  We prove
that the feedback capacity is given by
\begin{align}
C_\textsl{FB} &= \lim_{n\to\infty} C_{\textsl{FB},n} \notag\\
&= \limp_{n\to\infty} \sup_{p(x^n||y^{n-1})} 
\frac{1}{n} I(X^n \to Y^n), \label{eq:feedback2}
\end{align}
where the supremum is over all causally conditioned distributions
\[
p(x^n||y^{n-1}) = \prod_{i=1}^n p(x_i|x^{i-1}, y^{i-1}).
\]
We will combine the coding technique developed in the previous section
with the Shannon strategy for channels with side information, in
particular, Lemma~\ref{lemma:state2}.

That the limit in \eqref{eq:feedback2} is well-defined follows from the
superadditivity of $n C_{\textsl{FB},n}$.  Thus,
\[
C_\textsl{FB} = \limp_{n\to\infty} C_{\textsl{FB},n}
= \sup_{n \ge 1} C_{\textsl{FB},n}.
\]

The converse was proved by Massey~\cite[Theorem 3]{Massey1990}. We
repeat the proof here for completeness.  For any sequence of $(2^{nR},
n)$ codes with $P_e^{(n)}$, we have from Fano's inequality
\begin{align}
nR &\le I(W; Y^n) + n\eps_n \notag \\
&= \sum_{i=1}^n I(W; Y_i | Y^{i-1}) + n \eps_n \notag \\
&=  \sum_{i=1}^n I(X^i; Y_i | Y^{i-1}) + n \eps_n \label{eq:second-eq}\\
&= I(X^n \to Y^n) + n\eps_n,\notag
\end{align}
where $\eps_n \to 0$ as $n \to \infty$.  Here \eqref{eq:second-eq}
follows from the codebook structure $X_i(W, Y^{i-1})$ and the Markovity
$W \to (X^i, Y^{i-1}) \to Y_i$.

For the achievability, we show that there exists a sequence of codes
that achieves $C_{\textsl{FB},n}$ for each $n$.  As before, we assume
that the alphabets are finite.  In the light of
Lemma~\ref{lemma:state2}, it suffices to show that
\begin{equation}
\label{eq:c-fb}
C'_{\textsl{FB},n} = 
\max_{p(u^n), x_i = f(u_i, x^{i-1}, y^{i-1})} I(U^n; Y^n)
\end{equation}
is achievable, where the auxiliary random variables $U_i$ has the
cardinality bounded by $|\mathcal{U}_i| \le
|\mathcal{X}|^{i}|\mathcal{Y}|^{i-1}$, and the maximization is over
all joint distributions of the form
\[
p(u^n, x^n, y^n) 
= \Bigl(\prod_{i=1}^n
p(u_i) p(x_i | u_i, x^{i-1}, y^{i-1}) 
\Bigr)
p(y^n||x^n)
\]
with deterministic $p(x_i|u_i, x^{i-1}, y^{i-1})$, $i=1,\ldots,n$.

{\it Codebook generation and encoding.} Fix $n$ and let $p_i^*(u_i),$
$i=1,\ldots, n,$ and $f_i^*: (u_i, x^{i-1}, y^{i-1}) \mapsto x_i,$
$i=1,\ldots, n,$ achieve the maximum of \eqref{eq:c-fb}.  We will also
use the notation $p^*(u^n) = \prod_{i=1}^n p_i^*(u_i)$ and $f^*(u^n,
x^{n-1}, y^{n-1}) = (f_1^*(u_1),\ldots, f_n^*(u_n, x^{n-1},
y^{n-1}))$.

\begin{figure}[!t]
\begin{center}
\input{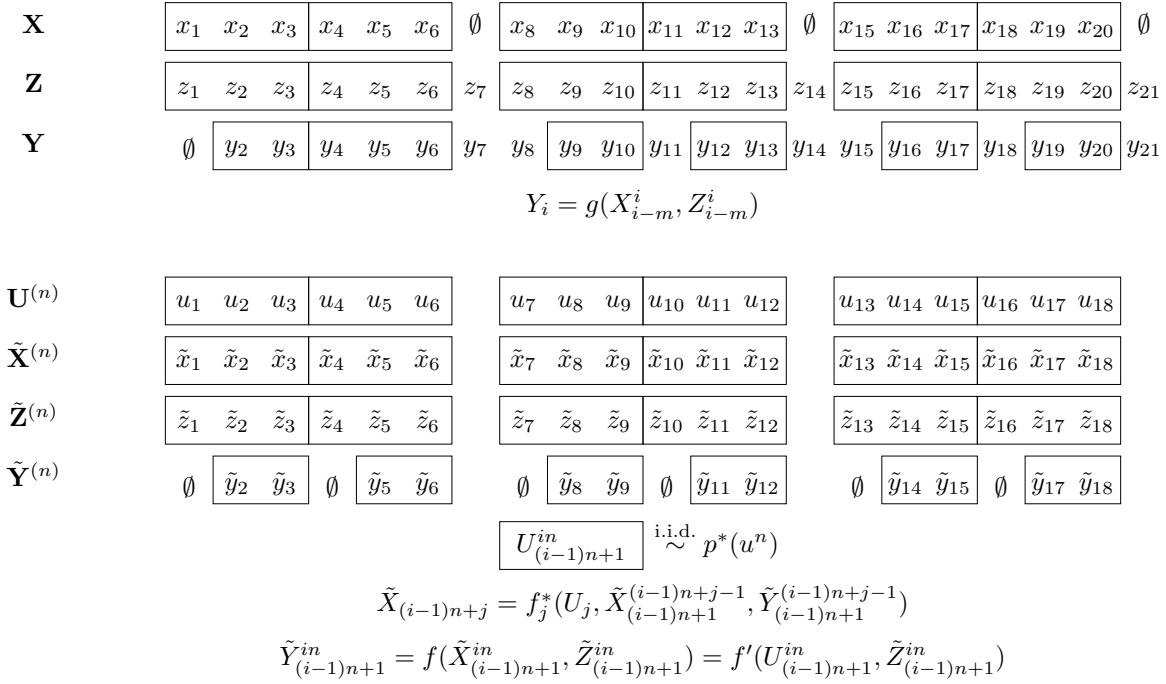}
\end{center}
\caption{Code, input, noise, and output sequences: $n = 3, L = 2, m =
1$.}
\label{fig:const5}
\end{figure}

For each $k = k(L,n) = Ln^2 + n,$ $L=1,2,\ldots,$ we generate a
$(2^{kR}, k)$ code $\{X_i(W, Y^{i-1})\}_{i=1}^k$ as summarized in
Figure~\ref{fig:const5}.  As before, $\tilde{\mathbf{X}}^{(n)},$
$\tilde{\mathbf{Y}}^{(n)}$, and $\tilde{\mathbf{Z}}^{(n)}$ are
respectively related to the underlying sequences $\mathbf{X},
\mathbf{Y}, \mathbf{Z}$ with every $(Ln+1)st$ symbol omitted.

For each $w \in \{1,2,\ldots,2^{kR}\}$, we generate a codeword
$\mathbf{U}^{(n)}(w) = {U}^{Ln^2}(w)$ of length $Ln$ on the $n$-letter
alphabet $\mathcal{U}_1 \times \cdots \times \mathcal{U}^n$
independently according to
\[
p(u^{Ln^2}) = \prod_{i=1}^{Ln} p^*(u_{(n-1)i + 1}^{ni}).
\]
This gives a $2^{kR} \times Ln$ codebook matrix with each entry
drawn \iid according to $p^*(u^n)$.

To communicate the message $W = w$, the transmitter
chooses the codeword $\mathbf{U}^{(n)}(w) = U^{Ln^2}(w)$ and sends
\[
\tilde{X}_{(i-1)n+j} = f_j^*(U_j(w), \tilde{X}_{(i-1)n+1}^{(i-1)n + j - 1}, 
\tilde{Y}_{(i-1)n+1}^{(i-1)n+j-1}),
\qquad i = 1,\ldots, Ln, \enspace j = 1,\ldots, n.
\]
Thus, the code function $X^n(w, Y^{n-1})$ utilizes the codeword
$\mathbf{U}^{(n)}$ and the channel feedback
$\tilde{\mathbf{Y}}^{(n)}$ only within the frame of $n$
transmissions (each box in Figure~\ref{fig:const5}).

{\it Decoding.}  Upon receiving $Y^k$, the receiver declares that the
message $\hat{W}$ was sent if there is a unique $\hat{W}$ such that
\[
(\mathbf{U}^{(n)}(\hat{W}), \tilde{\mathbf{Y}}^{(n)}) \in
A_\eps^{*(Ln)}(U^n,Y^n),
\]
that is, $(\mathbf{U}^{(n)}(\hat{W}), \tilde{\mathbf{Y}}^{(n)})$ is
jointly typical with respect to the joint distribution $p(u^n, y^n)$
specified by $p^*(u^n) p(z^n)$, $x_i = f^*_i(u_i, x^{i-1}, y^{i-1})$,
and the definition of the channel \eqref{eq:thm-channel-fb}.
Otherwise, an error is declared.

{\it Analysis of the probability of error.} We define the following
events:
\[
E_i = \{(\mathbf{U}^{(n)}(i), \tilde{\mathbf{Y}}^{(n)})
\in A_\eps^{*(Ln)}(U^n, Y^n)\},\qquad
i \in \{1,2,\ldots, 2^{kR} \}.
\]
As before, we assume $W = 1$ was sent.

From Lemma~\ref{lemma:joint-super}, $\mathbf{U}^{(n)}(1)$ and
$\mathbf{Z}^{(n)}$ are jointly typical with high probability for $L$
sufficiently large.  Furthermore, $\tilde{\mathbf{Y}}^{(n)}$ is an
$n$-letter blockwise function of $(\tilde{\mathbf{X}}^{(n)}(1),
\tilde{\mathbf{Z}}^{(n)})$, and thus of $(\mathbf{U}^{(n)}(1),
\tilde{\mathbf{Z}}^{(n)})$.  Therefore, the probability of the event
$E_1^c$ that the intended codeword $\mathbf{U}^{(n)}(1)$ is not
jointly typical with $\tilde{\mathbf{Y}}^{(n)}$ vanishes as $L \to
\infty$.

On the other hand, $\mathbf{U}^{(n)}(i),$ $i \ne 1,$ is generated
blockwise \iid $\sim p^*(u^n)$ independent of $\mathbf{Y}^{(n)}$.
Hence, from Lemma~\ref{lemma:joint-typicality}, the probability of
the event $E_i$ that $\mathbf{U}^{(n)}(i)$ is jointly typical with
$\mathbf{Y}^{(n)}$ is bounded by
\[
\Pr(E_i) \le 2^{-Ln (I(U^n; Y^n)-\delta)},\qquad \text{ for all } i \ne 1,
\]
where $\delta \to 0$ as $\eps \to 0$.  Consequently, we have
\begin{align*}
\Pr (\hat{W} \ne W) 
&\le \Pr(E_1^c) + \sum_{i=2}^{2^{kR}} \Pr(E_i) \\
&\le \eps + 2^{kR} 2^{-Ln (I(U^n; Y^n)-\delta)} \\
&\le 2 \eps
\end{align*}
if $L$ is sufficiently large and 
\[
kR = (Ln^2+n)R < Ln (I(U^n; Y^n) -
\delta).
\]
Thus by letting $L \to \infty$ and then $\eps\to 0$, we can achieve
any rate $R < C'_{\textsl{FB},n}$.

Finally by Lemma~\ref{lemma:state2}, this implies that we can achieve
\[
C_{\textsl{FB},n} = \max_{p(x^n||y^{n-1})} \frac{1}{n} I(X^n \to Y^n),
\]
which completes the proof of Theorem~\ref{thm:feedback}.

\section{Concluding Remarks}
\label{sec:conc}
Trading off generality off for transparency, we have focused on the
stationary channels of the form
\[
Y_n = f(X_{n-m}^n, Z_{n-m}^n)
\]
and presented a simple and constructive proof of the feedback coding
theorem.  The Shannon strategy (Lemma~\ref{lemma:state2}) has a
fundamental role in transforming the feedback coding problem into a
nonfeedback one, which is then solved by a scalable coding scheme of
constructing a long typical input-output sequence pair by
concatenating shorter nonergodic ones with appropriate phase shifts.

This two-stage approach can be applied to other channel models and
give a straightforward coding theorem.  For example, we can show that
the finite-state channel
\[
p(y_n,s_n|s_{n-1}, x_n) = p(y_n|s_{n-1},x_n) p(s_n|s_{n-1},x_n,y_n)
\]
with deterministic $p(s_n|s_{n-1},x_n,y_n)$ (but no assumption of
indecomposability) has the feedback capacity lower bounded by
\[
C_\textsl{FB} \ge \sup_{n\ge 1} \max_{p(x^n||y^{n-1})} \min_{s_0}
\frac{1}{n} I(X^n \to Y^n|s_0).
\]
This result was previously shown by Permuter {\it et
al.}~\cite[Section V]{Permuter--Weissman--Goldsmith2006} via a
generalization of Gallager's random coding exponent method for finite
state channels without feedback~\cite[Section 5.9]{Gallager1968}.
Here we sketch a simple alternative proof.

From a trivial modification of Lemma~\ref{lemma:state2}, the problem
reduces to showing that
\begin{equation}
\label{eq:fsc-det}
\max_{p(u^n),x_i = f(u_i, x^{i-1}, y^{i-1})} \min_{s_0} 
\frac{1}{n} I(U^n; Y^n|s_0)
\end{equation}
is achievable for each $n$.  But the given Shannon strategy
$(p^*(u^n), x^n = f^*(u^n, x^{n-1}, y^{n-1}))$ induces a new
time-invariant finite-state channel on the $n$-letter super alphabet
as $p(\mathbf{y}_k, \mathbf{s}_k | \mathbf{s}_{k-1}, \mathbf{u}_k)$.
Hence we can use Gallager's random coding exponent method directly to
achieve
\[
\limp_{k\to 1} \max_{p(\mathbf{u}^k)} \min_{\mathbf{s}_0} \frac{1}{k} 
I(\mathbf{U}^k; \mathbf{Y}^k|\mathbf{s}_0),
\]
which can be shown to be larger than our target
\[
\frac{1}{n} I(\mathbf{U}_1; \mathbf{Y}_1 | \mathbf{s}_0 ),
\]
because of the deterministic evolution of the state $S_n = f(S_{n-1},
X_n, Y_n)$.

We finally mention an important question that is not dealt with in
this paper.  Our characterization of the feedback capacity
\begin{equation}
\label{eq:fb-final}
C_\textsl{FB} = \lim_{n\to\infty} \max_{p(x^n||y^{n-1})} 
\frac{1}{n} I(X^n \to Y^n)
\end{equation}
or any similar multi-letter expressions are in general not computable
and do not provide much insight on the structure of the capacity
achieving coding scheme.  One may ask whether a stationary or even
Markov distribution is asymptotically optimal for the sequence of
maximizations in \eqref{eq:fb-final}.  This problem has been solved
for a few specific channel models such as certain classes of
finite-state channels~\cite{Chen--Berger2005,
Yang--Kavcic--Tatikonda2005, Permuter--Weissman--Goldsmith2006,
Permuter--Cuff--Van-Roy--Weissman2006} and stationary additive
Gaussian noise channels~\cite{Kim2006a, Kim2006c}, sometimes with
analytic expressions for the feedback capacity.  In this context, the
current development is just the first step toward the complete
characterization of the feedback capacity.

\section*{Acknowledgment} 
The author wishes to thank Tom Cover, Bob Gray, and Haim Permuter for
helpful discussions.

\def\cprime{$'$} \def\cprime{$'$} \def\cprime{$'$}


\begin{thebibliography}{10}
\providecommand{\url}[1]{#1}
\csname url@rmstyle\endcsname
\providecommand{\newblock}{\relax}
\providecommand{\bibinfo}[2]{#2}
\providecommand\BIBentrySTDinterwordspacing{\spaceskip=0pt\relax}
\providecommand\BIBentryALTinterwordstretchfactor{4}
\providecommand\BIBentryALTinterwordspacing{\spaceskip=\fontdimen2\font plus
\BIBentryALTinterwordstretchfactor\fontdimen3\font minus
  \fontdimen4\font\relax}
\providecommand\BIBforeignlanguage[2]{{%
\expandafter\ifx\csname l@#1\endcsname\relax
\typeout{** WARNING: IEEEtran.bst: No hyphenation pattern has been}%
\typeout{** loaded for the language `#1'. Using the pattern for}%
\typeout{** the default language instead.}%
\else
\language=\csname l@#1\endcsname
\fi
#2}}

\bibitem{Alajaji1995}
F.~Alajaji, ``Feedback does not increase the capacity of discrete channels with
  additive noise,'' \emph{{IEEE} Trans. Inf. Theory}, vol. IT-41, no.~2, pp.
  546--549, Mar. 1995.

\bibitem{Berger1971}
T.~Berger, \emph{Rate Distortion Theory}.\hskip 1em plus 0.5em minus
  0.4em\relax Englewood Cliffs, NJ: Prentice-Hall, 1971.

\bibitem{Blackwell1961}
D.~Blackwell, ``Information theory,'' in \emph{Modern Mathematics for the
  Engineer: Second Series}.\hskip 1em plus 0.5em minus 0.4em\relax New York:
  McGraw-Hill, 1961, pp. 182--193.

\bibitem{Brown1976}
J.~R. Brown, \emph{Ergodic Theory and Topological Dynamics}.\hskip 1em plus
  0.5em minus 0.4em\relax New York: Academic Press, 1976.

\bibitem{Caire--Shamai1999}
G.~Caire and S.~Shamai, ``On the capacity of some channels with channel state
  information,'' \emph{{IEEE} Trans. Inf. Theory}, vol. IT-45, no.~6, pp.
  2007--2019, 1999.

\bibitem{Chen--Berger2005}
J.~Chen and T.~Berger, ``The capacity of finite-state {M}arkov channels with
  feedback,'' \emph{{IEEE} Trans. Inf. Theory}, vol. IT-51, no.~3, pp.
  780--798, Mar. 2005.

\bibitem{Cover1975b}
T.~M. Cover, ``An achievable rate region for the broadcast channel,''
  \emph{{IEEE} Trans. Inf. Theory}, vol. IT-21, pp. 399--404, 1975.

\bibitem{Cover--Pombra1989}
T.~M. Cover and S.~Pombra, ``{G}aussian feedback capacity,'' \emph{{IEEE}
  Trans. Inf. Theory}, vol. IT-35, no.~1, pp. 37--43, Jan. 1989.

\bibitem{Cover--Thomas2006}
T.~M. Cover and J.~A. Thomas, \emph{Elements of Information Theory},
  2nd~ed.\hskip 1em plus 0.5em minus 0.4em\relax New York: Wiley, 2006.

\bibitem{Dobrushin1963}
R.~L. Dobrushin, ``General formulation of {S}hannon's main theorem in
  information theory,'' \emph{Uspkhi Mat. Nauk}, vol.~14, no.~6, pp. 3--104,
  1959, {E}nglish transl. in {\it Amer. Math. Soc. Transl.}, vol.~33, no.~2,
  pp.~323--438, 1963.

\bibitem{Feinstein1959}
A.~Feinstein, ``On the coding theorem and its converse for finite-memory
  channels,'' \emph{Information and Control}, vol.~2, pp. 25--44, 1959.

\bibitem{Feinstein1954}
------, ``A new basic theorem of information theory,'' \emph{{IRE} Trans. Inf.
  Theory}, vol. IT-4, pp. 2--22, 1954.

\bibitem{Forney1972}
G.~D. Forney, Jr., \emph{Information Theory}, unpublished course notes, 
{S}tanford {U}niversity, 1972.

\bibitem{Gallager1968}
R.~G. Gallager, \emph{Information Theory and Reliable Communication}.\hskip 1em
  plus 0.5em minus 0.4em\relax New York: Wiley, 1968.

\bibitem{Gallager1965}
------, ``A simple derivation of the coding theorem and some applications,''
  \emph{{IEEE} Trans. Inf. Theory}, vol. IT-11, pp. 3--18, 1965.

\bibitem{Gray1990}
R.~M. Gray, \emph{Entropy and Information Theory}.\hskip 1em plus 0.5em minus
  0.4em\relax New York: Springer-Verlag, 1990.

\bibitem{Gray--Ornstein1979}
R.~M. Gray and D.~S. Ornstein, ``Block coding for discrete stationary {$\bar
  d$}-continuous noisy channels,'' \emph{{IEEE} Trans. Inf. Theory}, vol.
  IT-25, no.~3, pp. 292--306, 1979.

\bibitem{Han2003}
T.~S. Han, \emph{Information-Spectrum Methods in Information Theory}.\hskip 1em
  plus 0.5em minus 0.4em\relax New York: Springer, 2003.

\bibitem{Khinchin1957}
A.~I. Khinchin, \emph{Mathematical Foundations of Information Theory}.\hskip
  1em plus 0.5em minus 0.4em\relax New York: Dover, 1957.

\bibitem{Kieffer1981a}
J.~C. Kieffer, ``Block coding for weakly continuous channels,'' \emph{{IEEE}
  Trans. Inf. Theory}, vol. IT-27, no.~6, pp. 721--727, 1981.

\bibitem{Kim2006a}
------, ``Feedback capacity of the first-order moving average {G}aussian
  channel,'' \emph{{IEEE} Trans. Inf. Theory}, vol. IT-52, no.~7, pp.
  3063--3079, 2006.

\bibitem{Kim2006c}
\BIBentryALTinterwordspacing
Y.-H. Kim, ``Feedback capacity of stationary {G}aussian channels,'' submitted
  to {\it IEEE Trans. Inf. Theory,} February 2006. [Online]. Available:
  \url{http://arxiv.org/abs/cs.IT/0602091/}
\BIBentrySTDinterwordspacing

\bibitem{Kramer1998}
G.~Kramer, \emph{Directed Information for Channels with Feedback}.\hskip 1em
  plus 0.5em minus 0.4em\relax Konstanz: Hartung-Gorre Verlag, 1998, {D}r. sc.
  thchn. Dissertation, Swiss Federal Institute of Technology (ETH) Zurich.

\bibitem{Kramer2003}
------, ``Capacity results for the discrete memoryless network,'' \emph{{IEEE}
  Trans. Inf. Theory}, vol. IT-49, no.~1, pp. 4--21, 2003.

\bibitem{Lapidoth--Telatar1998}
A.~Lapidoth and {\.I}.~E. Telatar, ``The compound channel capacity of a class
  of finite-state channels,'' \emph{{IEEE} Trans. Inf. Theory}, vol. IT-44,
  no.~3, pp. 973--983, 1998.

\bibitem{Massey1990}
J.~L. Massey, ``Causality, feedback, and directed information,'' in \emph{Proc.
  International Symposium on Information Theory and its Applications},
  Honolulu, Hawaii, Nov. 1990, pp. 303--305.

\bibitem{Massey--Massey2005}
J.~L. Massey and P.~C. Massey, ``Conservation of mutual and directed
  information,'' in \emph{Proc. International Symposium on Information Theory},
  Adelaide, Australia, Sept. 2005, pp. 157--158.

\bibitem{Nedoma1963}
J.~Nedoma, ``\"{U}ber die {E}rgodizit\"at und {$r$}-{E}rgodizit\"at
  station\"arer {W}ahrscheinlichkeitsmasse,'' \emph{Z.
  Wahrscheinlichkeitstheorie und Verw. Gebiete}, vol.~2, pp. 90--97, 1963.

\bibitem{Neuhoff--Shields1979}
D.~L. Neuhoff and P.~C. Shields, ``Channels with almost finite memory,''
  \emph{{IEEE} Trans. Inf. Theory}, vol. IT-25, no.~4, pp. 440--447, 1979.

\bibitem{Permuter--Cuff--Van-Roy--Weissman2006}
\BIBentryALTinterwordspacing
H.~Permuter, P.~Cuff, B.~Van~Roy, and T.~Weissman, ``Capacity of the trapdoor
  channel with feedback,'' submitted to \emph{IEEE Trans. Inform. Theory},
  2006. [Online]. Available: \url{http://arxiv.org/abs/cs.IT/0610047/}
\BIBentrySTDinterwordspacing

\bibitem{Permuter--Weissman--Goldsmith2006}
\BIBentryALTinterwordspacing
H.~Permuter, T.~Weissman, and A.~Goldsmith, ``Finite-state channels with
  time-invariant deterministic feedback,'' submitted to \emph{IEEE Trans.
  Inform. Theory}, 2006. [Online]. Available:
  \url{http://arxiv.org/abs/cs.IT/0608070/}
\BIBentrySTDinterwordspacing

\bibitem{Petersen1983}
K.~Petersen, \emph{Ergodic Theory}.\hskip 1em plus 0.5em minus 0.4em\relax
  Cambridge: Cambridge University Press, 1983.

\bibitem{Pinsker1964}
M.~S. Pinsker, \emph{Information and Information Stability of Random Variables
  and Processes}.\hskip 1em plus 0.5em minus 0.4em\relax San Francisco:
  Holden-Day, 1964.

\bibitem{Shannon1948}
C.~E. Shannon, ``A mathematical theory of communication,'' \emph{Bell System
  Tech. J.}, vol.~27, pp. 379--423, 623--656, 1948.

\bibitem{Shannon1956}
------, ``The zero error capacity of a noisy channel,'' \emph{{IRE} Trans. Inf.
  Theory}, vol. IT-2, no.~3, pp. 8--19, Sept. 1956.

\bibitem{Shannon1958a}
------, ``Channels with side information at the transmitter,'' \emph{IBM J.
  Res. Develop.}, vol.~2, pp. 289--293, 1958.

\bibitem{Tatikonda2000}
S.~Tatikonda, ``Control under communication constraints,'' Ph.D. Thesis,
  Massachusetts Institute of Technology, Sept. 2000.

\bibitem{Verdu--Han1994}
S.~Verd{\'u} and T.~S. Han, ``A general formula for channel capacity,''
  \emph{{IEEE} Trans. Inf. Theory}, vol. IT-40, no.~4, pp. 1147--1157, July
  1994.

\bibitem{Wolfowitz1978}
J.~Wolfowitz, \emph{Coding Theorems of Information Theory}, 3rd~ed.\hskip 1em
  plus 0.5em minus 0.4em\relax Berlin: Springer-Verlag, 1978.

\bibitem{Yang--Kavcic--Tatikonda2005}
S.~Yang, A.~Kav{\v{c}}i{\'c}, and S.~Tatikonda, ``Feedback capacity of
  finite-state machine channels,'' \emph{{IEEE} Trans. Inf. Theory}, vol.
  IT-51, no.~3, pp. 799--810, Mar. 2005.

\end{thebibliography}

\end{document}